\documentclass[intlimits,twoside,a4paper]{article}

\usepackage[cp1251]{inputenc}
\usepackage{xcolor}
\usepackage[eqsecnum]{cmpj3}
\usepackage{bm}
\DeclareMathOperator{\sgn}{sgn}
\issue{2020}{23}{2}{23003}
\doinumber{10.5488/CMP.23.23003}

\title[Generalized diffusion equation with nonlocality of space-time. Memory function modelling] %
{Generalized diffusion equation with nonlocality of space-time. Memory function modelling}
\author[P.P. Kostrobij, B.M. Markovych, M.V. Tokarchuk]{P.P. Kostrobij\refaddr{label1}, B.M. Markovych\refaddr{label1}, M.V. Tokarchuk\refaddr{label2,label1}}
\addresses{
\addr{label1} Lviv Polytechnic National University, 12 S. Bandera St., 79013 Lviv, Ukraine
\addr{label2} Institute for Condensed Matter Physics of the National
Academy of Sciences of Ukraine,\\ 1 Svientsitskii St., 79011 Lviv,
Ukraine
}

\newcommand{\DDD}[3]{\raisebox{-.6ex}{\scriptsize #1}\mathrm{D}_{#2}^{#3}}

\date{Received January 31, 2020, in final form April 5, 2020}
\begin{document}

\maketitle

\begin{abstract}
  We presented a general approach for obtaining the generalized transport equations with fractional derivatives by using the Liouville equation with fractional derivatives for a system of classical particles and Zubarev's nonequilibrium statistical operator (NSO) method within Gibbs statistics.
  The new non-Markovian diffusion equations of ions in spatially heterogeneous environment with fractal structure and generalized Cattaneo-Maxwell diffusion equation with taking into account the space-time nonlocality are obtained.
  Dispersion relations are found for the Cattaneo-Maxwell diffusion equation with taking into account the space-time nonlocality in fractional derivatives.
  The frequency spectrum, phase and group velocities are calculated.
  It is shown that it has a wave behaviour with discontinuities,
  which are also manifested in the behaviour of the phase velocity.
\end{abstract}

\section{Introduction}

 Studies of nonequilibrium processes with spatio-temporal nonlocality are relevant in the statistical physics of soft matter.

 One of the important problems in the theory of nonequilibrium processes of interacting particles is the calculation of memory functions (transfer kernels) in transport equations in a wide region of spatio-temporal dependence,
 including the region of anomalous behaviour,
 in particular, sub-, superdiffusions,
 which are experimentally realized in condensed systems.

 Mathematical modelling of diffusion (sub-, superdiffusion)  transfer processes in porous and complex nano-structured (with characteristic fractality) systems requires the use of transfer equations with significant spatial inhomogeneity and temporal memory.
 An actual problem for description of nonequilibrium processes in complex systems is the construction of generalized diffusion and wave equations~\cite{Sandev20151006,Sandev2018015201,Sandev2019} using fractional integrals and derivatives.

 Relevance and research methods in this field are discussed in \cite{Kostrobij201963} (and references therein).

 The dispersion of heat waves in a dissipative environment using the Cattaneo-Maxwell heat diffusion equation with fractional derivatives is investigated in \cite{Giusti2018013506}.
  Based on this equation,
 the frequency spectrum, phase and group velocities of propagation of heat waves in a dissipative environment are investigated.

 In our works~\cite{Kostrobij201963,Kostrobij2016093301,Kostrobij2016163,Glushak201857,Kostrobij201875,Kostrobij201958,Grygorchak2015e,
 Kostrobij2015154,Grygorchak2017185501,Kostrobij20184099,Kostrobij2019289}, a statistical approach to obtain generalized spatio-temporal nonlocal transfer equations was developed by using the Zubarev nonequilibrium statistical operator (NSO) method~\cite{Zubarev19811509,Zubarev20021and2,Markiv2011785} and the Liouville equation with fractional derivatives~\cite{Tarasov2010,Tarasov2004123}.
 In particular,
 the generalized diffusion equations of Cattaneo~\cite{Kostrobij2016163,Kostrobij201875,Kostrobij201963},
 Cattaneo-Maxwell~\cite{Kostrobij201958} and electrodiffusion~\cite{Grygorchak2015e,Kostrobij2015154,Grygorchak2017185501,Kostrobij20184099},
 kinetic equations~\cite{Kostrobij2019289} with spatio-temporal fractional derivatives were obtained.

 In the second section,
 we present a statistical approach to construct generalized transfer equations using the NSO method and the Liouville equation with fractional derivatives.
 In the third section,
 a generalized diffusion equation with fractional derivatives is obtained based on this approach.
 Generalized Cattaneo-Maxwell diffusion equation with fractional derivatives will be obtained by modelling the memory function and using fractional calculus methods.
 In the fourth section,
 model calculations of the frequency spectrum of the Cattaneo-Maxwell diffusion equation with fractional derivatives are performed.

\section{Liouville equation with fractional derivatives for classical system of particles}

 We use the Liouville equation with fractional derivatives obtained by Tarasov in \cite{Tarasov2004123,Tarasov2010}
 for a nonequilibrium particle function $\rho(x^{N};t)$ of a classical system
 \begin{equation}\label{eq:2.1}
   \frac{\partial }{\partial t}\rho(x^{N};t)+\ri L_{\alpha}\rho(x^{N};t)=0,
 \end{equation}
 where $\ri L_{\alpha}$ is the Liouville operator with the fractional derivatives,
 \begin{equation}\label{eq:2.2}
  \ri L_{\alpha}\rho(x^{N};t)
  =
  \sum^{N}_{j=1}
  \left[
  \mathrm{D}^{\alpha}_{\vec{p}_{j}}H(\vec{r},\vec{p})\mathrm{D}^{\alpha}_{\vec{r}_{j}}
  -
  \mathrm{D}^{\alpha}_{\vec{r}_{j}}H(\vec{r},\vec{p})\mathrm{D}^{\alpha}_{\vec{p}_{j}}\right]%\nonumber\\
  \rho(x^{N};t),
 \end{equation}
${x^{N}=x_{1},\ldots,x_{N}}$, 
 ${x_{j}=\{\vec{r}_{j}, \vec{p}_{j}\}}$ are dimensionless generalized coordinates,
 ${\vec{r}_{j}=(r_{j1},\ldots,r_{jm})}$,
 and generalized momentum,
 ${\vec{p}_{j}=(p_{j1},\ldots,p_{jm})}$,~\cite{Tarasov2006341} of $j$-th particle
 in the phase space with a fractional differential volume  element~\cite{Kathleen20012203,Tarasov2004123} $\rd^{\alpha}V=\rd^{\alpha}x_{1}\ldots \rd^{\alpha}x_{N}$.
 Here,
 $m={Mr_{0}}/{p_{0}t_{0}}$,
 $M$ is the mass of particle,
 $r_{0}$ is a characteristic scale in the configuration space,
 $p_{0}$  is a characteristic momentum,
 and $t_{0}$ is a characteristic time.
 $\rd^{\alpha}$ is a fractional differential~\cite{Kathleen20012203} that is defined by
 $\rd^{\alpha} f(x)=\sum^{2N}_{j=1}\mathrm{D}^{\alpha}_{x_{j}}f(x)(\rd x_{j})^{\alpha}$,
  where  $(\rd x_{j})^{\alpha}:=\sgn(\rd x_{j})|\rd x_{j}|^{\alpha}$~\cite{Tarasov20082756,Tarasov2011e},
 $ %\begin{equation}\label{eq:2.2}
  \mathrm{D}^{\alpha}_{x} f(x)=\frac{1}{\Gamma (n-\alpha)} \int^{x}_{0} \frac{f^{(n)}(z)}{(x-z)^{\alpha +1-n}} \rd z
 $ %\end{equation}
 is the Caputo fractional derivative~\cite{Mainardi1997,Caputo1971134,Oldham2006,Samko1993},
 ${n-1<\alpha<n}$,
 $f^{(n)}(z)=\frac{\rd^{n}}{\rd z^{n}}f(z)$ with the properties $\mathrm{D}^{\alpha}_{x_{j}}1=0$ and $\mathrm{D}^{\alpha}_{x_{j}}x_{l}=0$, $(j\neq l)$.
 A solution of the Liouville equation (\ref{eq:2.1}) will be found with the Zubarev nonequilibrium statistical operator method~\cite{Zubarev19811509,Zubarev20021and2}.
 After choosing parameters of the reduced description,
 taking into account the projections, we present the nonequilibrium particle function $\rho(x^{N};t)$
 (as a solution of the Liouville equation) in the general form
 \begin{equation}\label{eq:2.6}
   \rho(x^{N};t)
   =
   \rho_{\text{rel}}(x^{N};t)
   -
   \int^{t}_{-\infty}\re^{\varepsilon(t'-t)}T(t,t')%\nonumber\\
   \left[1-P_\text{rel}(t')\right]\ri L_{\alpha}\rho_{\text{rel}}(x^{N};t')\rd t',
 \end{equation}
 where
 $T(t,t')=\exp_{+}\lbrace -\int^{t}_{t'}[1-P_{\text{rel}}(t')]\ri L_{\alpha}\rd t'\rbrace $
 is the evolution operator in time containing the projection,
 $\exp_+$ is ordered exponential,
 ${\varepsilon\to+0}$ after taking the thermodynamic limit,
 $P_{\text{rel}}(t')$ is the generalized Kawasaki-Gunton projection operator depending on a structure of the relevant statistical operator (distribution function),
 $\rho_{\text{rel}}(x^{N};t')$.
 By using the Zubarev nonequilibrium statistical operator method~\cite{Zubarev19811509,Zubarev20021and2},
 $\rho_{\text{rel}}(x^{N};t')$ will be found from the extremum of the Gibbs entropy
 at fixed values of the observed values $\langle \hat{P}_{n}(x)\rangle^{t}_{\alpha}$,
 taking into account the normalization condition $\langle 1 \rangle^{t}_{\alpha, \text{rel}}=1$,
 where the nonequilibrium average values are found, respectively~\cite{Tarasov2010},
 \begin{equation}\label{eq:2.7}
  \langle \hat{P}_{n} (x)\rangle^{t}_{\alpha}=\hat{I}^{\alpha}(1,\ldots,N)\hat{T}(1,\ldots,N)\hat{P}_{n}\rho(x^{N};t).
 \end{equation}
 The operator $\hat{I}^{\alpha}(1,\ldots,N)$ has the following form for a system of $N$ particles
  $\hat{I}^{\alpha}(1,\ldots,N)
  =\hat{I}^{\alpha}(1)\ldots$ $\hat{I}^{\alpha}(N)$,
    $\hat{I}^{\alpha}(j)=\hat{I}^{\alpha}(\vec{r}_{j})\hat{I}^{\alpha}(\vec{p}_{j})$
  and defines the operation of integration
 $ %\begin{equation}\label{eq:2.8}
  \hat{I}^{\alpha}(x)f(x)=\int^{\infty}_{-\infty}f(x)\rd\mu_{\alpha}(x),
 $
 $
  \rd\mu_{\alpha}(x)=\frac{|x|^{\alpha}}{\Gamma (\alpha)}\rd x.
 $ %\end{equation}
 The operator $\hat{T}(1,\ldots,N)=\hat{T}(1)\ldots\hat{T}(N)$ defines the operation
 $
  \hat{T}(x_{j})f(x_{j})={1}/{2}[f(\ldots,x'_{j}-x_{j},\ldots)+f(\ldots,x'_{j}+x_{j},\ldots)] .
 $
 Accordingly,
 the average value,
 which is calculated with the relevant distribution function,
 is defined as
  $\langle (\ldots) \rangle^{t}_{\alpha, \text{rel}}=\hat{I}^{\alpha}(1,\ldots,N)\hat{T}(1,\ldots,N)(\ldots)\rho_{\text{rel}}(x^{N};t)$.

 According to \cite{Zubarev19811509,Zubarev20021and2},
 from the extremum of the Gibbs entropy  functional
 \begin{align*}
  L_{R}(\rho')&=-\hat{I}^{\alpha}(1,\ldots,N)\hat{T}(1,\ldots,N)\rho'(t) \ln\rho'(t)
  -\gamma\hat{I}^{\alpha}(1,\ldots,N)\hat{T}(1,\ldots,N)\rho'(t)\\
  &-\sum_{n}\int \rd\mu_{\alpha}(x)F_{n}(x;t)\hat{I}^{\alpha}(1,\ldots,N)\hat{T}(1,\ldots,N)\hat{P}_{n}(x)\rho'(t)
 \end{align*}
 at fixed values of the observed values $\langle \hat{P}_{n}(x)\rangle^{t}_{\alpha}$
 and
 the normalization condition $\hat{I}^{\alpha}(1,\ldots,N)\hat{T}(1,\ldots,N)$ $\times\rho'(t)=1$,
 the relevant distribution function takes the form
 \begin{equation}\label{eq:2.9}
  \rho_{\text{rel}}(t)
  =
  \frac{1}{Z_{\text{G}}(t)}
  \exp \left\lbrace -  \beta\left[ 
  H %\nonumber\\
  -\sum_{n}\int \rd\mu_{\alpha}(x)F_{n}(x;t) \hat{P}_{n}(x)
  \right] 
  \right\rbrace ,
 \end{equation}
 where
 $Z_{\text{G}}(t)$ is the partition function of the Gibbs distribution,
 which is determined from the normalization condition and has the form:
 $ %\begin{equation}\label{eq:2.10}
   Z_{\text{G}}(t)=\hat{I}^{\alpha}(1,\ldots,N)\hat{T}(1,\ldots,N)$ $
   \exp \lbrace -  \beta[
  H %\nonumber\\
  -\sum_{n}\int \rd\mu_{\alpha}(x)F_{n}(x;t) \hat{P}_{n}(x)
  ]\rbrace .
 $ %\end{equation}
 The Lagrangian multiplier $\gamma$ is determined by the normalization condition $\hat{I}^{\alpha}(1,\ldots,N)\hat{T}(1,\ldots,N)$ $\times\rho'(t)=1$.
 The parameters $F_{n}(x;t)$ are determined from the self-consistency conditions
 $ %\begin{equation}\label{eq:2.11}
   \langle \hat{P}_{n}(x) \rangle^{t}_{\alpha}= \langle \hat{P}_{n}(x) \rangle^{t}_{\alpha, \text{rel}}\,.
 $ %\end{equation}

 In the general case of the parameters $\langle \hat{P}_{n} (x)\rangle _{\alpha }^{t}$ of the reduced description of nonequilibrium processes according to \eqref{eq:2.6} and \eqref{eq:2.9},
 we get the nonequilibrium statistical operator in the form
 \begin{equation} \label{GrindEQ__12_}
  \rho (t)=\rho_{\text{rel}} (t)
  +\sum _{n}\int \rd\mu _{\alpha } (x  )
  \int _{-\infty }^{t} \re^{\varepsilon (t'-t)} T(t,t')I_{n} (x;t')\rho_{\text{rel}} (t')\beta F_{n} (x;t')\rd t',
 \end{equation}
 where $I_{n} (x;t')=[1-P(t)]\ri L_{\alpha } \hat{P}_{n} (x)$
  are the generalized flows,
 $P(t)$ is the Mori projection operator~\cite{Kostrobij2016093301}.

 By using the nonequilibrium statistical operator~\eqref{GrindEQ__12_},
 we get the generalized transport equation for the parameters $\langle \hat{P}_{n} (x)\rangle _{\alpha }^{t}$ of the reduced description
 \begin{equation} \label{GrindEQ__14_}
   \frac{\partial }{\partial t}\langle \hat{P}_{n} (x)\rangle _{\alpha }^{t} =\langle iL_{\alpha } \hat{P}_{n} (x)\rangle _{\alpha ,\text{rel}}^{t}
   +\sum _{n'}\int \rd\mu _{\alpha } (x'  )\int \limits_{-\infty }^{t}\re^{\varepsilon (t'-t)} \varphi_{P_{n} P_{n'}} (x,x';t,t')\beta F_{n'} (x';t')\rd t',
 \end{equation}
 where
 $ %\begin{equation} \label{GrindEQ__15_}
  \varphi_{P_{n} P_{n'} } (x,x';t,t')=\hat{I}^{\alpha } (1,\ldots,N)\hat{T}(1,\ldots,N)
  [I_{n} (x;t)T(t,t')I_{n'} (x';t')\rho_{\text{rel}} (x^{N} ;t')]
 $ %\end{equation}
 are the generalized transport kernels (the memory functions),
 which describe dissipative processes in the system.
 To demonstrate the structure of the transport equations~\eqref{GrindEQ__14_} and the transport kernels, %~\eqref{GrindEQ__15_},
 we will consider, for example, diffusion processes.
 In the next section,
 we obtain generalized transport equations with fractional derivatives and consider a concrete example of diffusion processes of the particle in non-homogeneous media.

 \section{Generalized diffusion equations with fractional derivatives}

 One of the main parameters for the reduced description of the diffusion processes of the particles in non-homogeneous media with fractal structure is the nonequilibrium density of the particle numbers,
 $\langle \hat{P}_{n}(x) \rangle^{t}_{\alpha}$:
 $n (\vec{r};t)=\langle \hat{n} (\vec{r})\rangle _{\alpha }^{t}$,
 where $\hat{n} (\vec{r})=\sum _{j=1}^{N } \delta (\vec{r}-\vec{r}_{j} )$ is the microscopic density of the particles.
 The corresponding generalized diffusion equation for $n (\vec{r}; t)$ can be obtained based on \eqref{eq:2.9}--\eqref{GrindEQ__14_}
 \begin{equation} \label{01}
    \frac{\partial }{\partial t} \left\langle \hat{n} (\vec{r})\right\rangle _{\alpha }^{t} =
    \mathrm{D}^{\alpha}_{\vec{r}} \cdot  \int  \rd\mu _{\alpha '} (\vec{r}' )
    \int \limits_{-\infty }^{t}\re^{\varepsilon (t'-t)} D^{\alpha\alpha'} \!\!(\vec{r},\vec{r}' ;t,t' )\cdot \mathrm{D}^{\alpha'}_{\vec{r}'} \beta \nu   (\vec{r}';t')\rd t' ,
 \end{equation}
 where
 $ %\begin{equation} \label{02}
  D^{\alpha\alpha'} (\vec{r},\vec{r}';t,t')=\langle \hat{\vec{v}}^{\alpha} (\vec{r})T(t,t')\hat{\vec{v}}^{\alpha'} (\vec{r}')\rangle _{\alpha ,\text{rel}}^{t}
 $ %\end{equation}
 is the generalized coefficient  diffusion of the particles within the Gibbs statistics.
 The averaging in $D^{\alpha\alpha'} (\vec{r},\vec{r}';t,t')$ % Eq.~\eqref{02}
 is performed with the power-law Gibbs distribution,
 $ %\begin{equation} \label{03}
  \rho_{\text{rel}} (t)=\frac{1}{Z_{\text G} (t)} \exp \lbrace - \beta [H- \int  \rd\mu _{\alpha } (\vec{r})\nu   (\vec{r};t)\hat{n} (\vec{r})]\rbrace  ,
 $ %\end{equation}
 where
 $ % \begin{equation} \label{04}
  Z_{\text G} (t)=\hat{I}^{\alpha } (1,\ldots,N)\hat{T}(1,\ldots,N)$ $\times\exp \{ - \beta [H- \int  \rd\mu _{\alpha } (\vec{r})\nu  (\vec{r};t)\hat{n} (\vec{r})]\}
 $ %\end{equation}
 is  the partition function of the relevant distribution function,
 $H$ is a Hamiltonian of the system.
 Parameter $\nu  (\vec{r};t)$
 is the chemical potential of the particles,
 which is determined from the self-consistency condition,
 \begin{equation} \label{05}
  \left\langle \hat{n} (\vec{r})\right\rangle _{\alpha }^{t} =\left\langle \hat{n} (\vec{r})\right\rangle _{\alpha ,\text{rel}}^{t} .
 \end{equation}
 $\beta ={1}/{k_{{\rm B}} T} $  ($k_{{\rm B}} $ is the Boltzmann constant),
 $T$ is the equilibrium value of temperature,
 $\hat{\vec{v}}^{\alpha} (\vec{r})=\sum _{j=1}^{N} \vec{v}^{\alpha}_{j} \delta (\vec{r}-\vec{r}_{j} )$
 is the microscopic flux density of the particles.

 In the Markov approximation,
 the generalized coefficient of  diffusion in time and space has the form
 $D^{\alpha\alpha'} (\vec{r},\vec{r}';t,t')\approx D \,\delta (t-t')\delta (\vec{r}-\vec{r}')\delta_{\alpha\alpha'}$.
 By excluding the parameter $\nu  (\vec{r}';t')$ via the self-consistency condition,
 we obtain the diffusion equation with fractional derivatives from \eqref{01}
 \begin{equation} \label{GrindEQ__26_}
  \frac{\partial }{\partial t} \left\langle \hat{n} (\vec{r})\right\rangle _{\alpha }^{t} =\sum _{b} D \,\mathrm{D}^{2\alpha}_{r} \nu (\vec{r};t').
 \end{equation}

 The generalized diffusion equation takes into account spatial nonlocality of the system and memory effects in the generalized coefficient of
  diffusion $D^{\alpha\alpha'} (\vec{r},\vec{r}';t,t')$ within the Gibbs statistics.
  To show the multifractal time in the generalized diffusion equation,
 we use the following approach for the generalized coefficient of particle diffusion
 \begin{equation}
   D^{\alpha\alpha'}\!(\vec r, \vec r';t,t')=W(t,t')\bar{D}^{\alpha\alpha'}\!\!(\vec r, \vec r'),
  \end{equation}
 where $W(t,t')$ can be defined as the time memory function.
 In view of this,
 \eqref {01} can be represented as
 \begin{equation}\label{eq:2.191}
   \frac{\partial}{\partial t}\left\langle\hat n(\vec r)\right\rangle^{t}_{\alpha}
   =\int_{-\infty}^{t}\!\!\re^{\varepsilon(t'-t)}W(t,t')\Psi(\vec r;t')\rd t',
 \end{equation}
 where
 $ %\begin{equation}\label{eq:2.192}
   \Psi(\vec r;t')
   =\int\! \rd\mu_{\alpha'}(\vec r')D^{\alpha}_{\vec{r}}\cdot \bar{D}^{\alpha\alpha'}(\vec r, \vec r')\cdot
  D^{\alpha'}_{\vec{r}'}\beta\nu(\vec r';t').
 $ %\end{equation}

 Further we apply the Fourier transform to \eqref{eq:2.191},
 and, as a result,  in frequency representation we get
 \begin{equation}\label{eq:2.193}
  \ri\omega n(\vec r;\omega) =W(\omega)\Psi(\vec r;\omega).
 \end{equation}

 We can represent the frequency dependence of the memory function  in the following form
 \begin{equation}\label{eq:2.194}
   W(\omega)=\frac{(\ri\omega)^{1-\xi}}{1+(\ri\omega \tau)^{\xi}}\, ,\quad  0<\xi \leqslant 1,
 \end{equation}
 where the introduced relaxation time $\tau$ characterizes  the particle transport processes in the system.
 Then,~\eqref{eq:2.193} can be represented as
 \begin{equation}\label{eq:2.195}
  \left[  1+(\ri\omega \tau)^{\xi}\right]\ri\omega n(\vec r;\omega) =(\ri\omega)^{1-\xi}\Psi(\vec r;\omega).
 \end{equation}

 Further we use the Fourier transform to fractional derivatives of functions
 \begin{equation}\label{eq:2.1096}
   L\big(\DDD{0}{t}{1-\xi}f(t);\ri\omega\big)=(\ri\omega)^{1-\xi} L(f(t);\ri\omega),
 \end{equation}
 where $\DDD{0}{t}{1-\xi}f(t)=\frac{1}{\Gamma(\xi)}\frac{\rd}{\rd t}\int^{t}_{0}
 \frac{f(\tau)}{(t-\tau)^{1-\xi}}\rd\tau$  is the Riemann-Liouville fractional derivative.
 By using it,
 the inverse transformation of \eqref{eq:2.195} to time representation yields the Cattaneo-Maxwell generalized diffusion equation
 with  taking into account spatial fractality,
  in the expanded form
 \begin{equation}\label{eq:2.1960}
   \DDD{0}{t}{2\xi} n(\vec r;t)\tau^{\xi} +\DDD{0}{t}{\xi} n(\vec r;t) =\int\! \rd\mu_{\alpha'}(\vec r')\mathrm{D}_{r}^{\alpha}\cdot \bar{D}(\vec r, \vec r')\cdot
  \mathrm{D}_{r'}^{\alpha'}\beta\nu(\vec r';t),
 \end{equation}
 which is the new Cattaneo-Maxwell generalized equation within the Gibbs statistics with  time and spatial nonlocality.
 Equation \eqref{eq:2.1960} contains significant spatial heterogeneity in $\bar{D}^{\alpha\alpha'}(\vec r, \vec r')$.
 If we neglect spatial heterogeneity,
 $ %\begin{equation}\label{eq:2.197}
     \bar{D}^{\alpha\alpha'}(\vec r, \vec r')= \bar{D}\,\delta (\vec r - \vec r')\delta_{\alpha\alpha'},
 $ %\end{equation}
 we get the Cattaneo-Maxwell diffusion equation with  space-time  nonlocality
 and constant coefficients of  diffusion within the Gibbs statistics
 \begin{equation}\label{eq:2.393}
  \DDD{0}{t}{2\xi} n(\vec r;t) \tau^{\xi} +\DDD{0}{t}{\xi} n(\vec r;t) =\bar{D}\,\mathrm{D}_{r}^{2\alpha}
    \beta \nu(\vec r;t).
 \end{equation}

 \section{Dispersion relation for the time-space-fractional Cattaneo-Maxwell diffusion equation}

 Using the self-consistent condition~\eqref{05} and the approved approximations,
 \eqref{eq:2.393} can be written as
 \begin{equation}\label{eq:2.40}
  \DDD{0}{t}{2\xi} n(\vec r;t) \tau^{\xi} +\DDD{0}{t}{\xi} n(\vec r;t) -\bar{D}'\,\mathrm{D}_{r}^{2\alpha} n(\vec r;t)=0,
 \end{equation}
 where $\bar{D}'$ is the renormalized diffusion coefficient.
 For simplicity,
 we consider the one-dimensional case and a solution of \eqref{eq:2.40} will be sought in the form of the plane wave,
 $n(x;t)\sim \re^{-\ri\omega t+\ri kx}$.
 Then, we get
% \begin{equation}\label{eq:3.2}
%  \big(\tau^{\xi}(-i\omega)^{2\xi}  + (-i\omega)^{\xi}\big)n(k;\omega)-\bar{D}'(ik)^{2\alpha}n(k;\omega)=0
% \end{equation}
 the corresponding frequency spectrum,
 $ %\begin{equation}\label{eq:3.3}
 \tau^{\xi}(-\ri\omega)^{2\xi}  + (-\ri\omega)^{\xi}-\bar{D}'(\ri k)^{2\alpha}=0.
 $ % \end{equation}

 Let us consider the case~$\alpha=1$ and $0<\xi<1$.
A similar problem is solved in \cite{Giusti2018013506} while investigating the dispersion relations for the Cattaneo-Maxwell heat transfer equation with fractional derivatives.
 At $\alpha=1$,
 we get the equation
 $ %\begin{equation}\label{eq:3.4}
  \tau^{\xi}(-\ri\omega)^{2\xi}  + (-\ri\omega)^{\xi}+\bar{D}'k^{2}=0,
 $ %\end{equation}
 the solution of which has the form:
 $ %\begin{equation}\label{eq:3.5}
  (-\ri\omega)^{\xi} =[-1\pm\sqrt{1-(k/k_0)^{2}}]/(2\tau^{\xi}),
 $ %\end{equation}
 where $k_0=1/\sqrt{4\tau^{\xi}\bar{D}'}$.
 Next, we find imaginary and real parts of the frequency by putting $\omega(k)=\omega_\text{r}(k)+\ri\omega_\text{i}(k)=|\omega(k)|\re^{\ri\theta(k)}$,
 $|\omega|>0$,
 $-\piup<\theta\leqslant\piup$.
 Then, we get
 \begin{equation*}
   \omega_\text{r}(k)=
   \left\{
     \begin{array}{ll}
       -\frac{1}{2^{1/\xi}\tau}\left[{1\pm\sqrt{1-(k/k_0)^{2}}}\right]^{{1}/{\xi}}\sin\frac{\piup}{\xi}\,, &
                      0\leqslant k \leqslant k_0\,,\\
       \mp \frac1\tau\left(\frac{k}{2k_0}\right)^{{1}/{\xi}}
                     \sin\left[-\frac\piup\xi+\frac{1}{\xi}\arctan \sqrt{(k/k_0)^{2}-1}\right],  & k > k_0\,,
     \end{array}
   \right.
 \end{equation*}
 \begin{equation*}
   \omega_\text{i}(k)=
   \left\{
     \begin{array}{ll}
       \frac{1}{2^{1/\xi}\tau}\left[{1\mp\sqrt{1-(k/k_0)^{2}}}\right]^{{1}/{\xi}}\cos\frac{\piup}{\xi}\,, &
                      0\leqslant k \leqslant k_0\,,\\
       \frac1\tau\left(\frac{k}{2k_0}\right)^{{1}/{\xi}}
                     \cos\left[-\frac\piup\xi+\frac{1}{\xi}\arctan \sqrt{(k/k_0)^{2}-1}\right],  & k > k_0.
     \end{array}
   \right.
 \end{equation*}

 According to definitions of the phase $v_\text{p}(k)$ and group velocities $v_\text{g}(k)$:
 $ %\begin{equation}\label{eq:3.11}
   v_\text{p}(k)=\frac{\omega_\text{r}(k)}{k},
 $
 $
   v_\text{g}(k)=\frac{\partial }{\partial k}\omega_\text{r}(k),
 $ %\end{equation}
 we obtain the following expressions for $v_\text{p}(k)$ and $v_\text{g}(k)$:
 \begin{equation*}
   v_\text{p}(k)=
   \left\{
     \begin{array}{ll}
       -\frac{1}{2^{1/\xi}\tau k}\left[{1\pm\sqrt{1-(k/k_0)^{2}}}\right]^{{1}/{\xi}}\sin\frac{\piup}{\xi}\,, &
                      0\leqslant k \leqslant k_0\,,\\
       \mp \frac1{\tau k}\left(\frac{k}{2k_0}\right)^{{1}/{\xi}}
                     \sin\left[-\frac\piup\xi+\frac{1}{\xi}\arctan \sqrt{(k/k_0)^{2}-1}\right],  & k > k_0\,,
     \end{array}
   \right.
 \end{equation*}
\begin{equation*}
v_\text{g}(k)=
\left\{
\begin{array}{ll}
\pm \frac{1}{2^{1/\xi}\tau \xi} \frac{k/k_0^2}{\sqrt{1-(k/k_0)^{2}}}
\left[{1\pm\sqrt{1-(k/k_0)^{2}}}\right]^{\frac{1-\xi}{\xi}}\sin\frac{\pi}{\xi}\,, &
	0\leqslant k \leqslant k_0\,,\\
\mp \frac{1}{\xi\tau k} \left({\frac{k}{2k_0}}\right)^{{1}/{\xi}}
\Big\{ \sin\left[-\frac\pi\xi+\frac{1}{\xi}\arctan \sqrt{(k/k_0)^{2}-1}\right]  \\
\qquad\qquad+\frac{1}{\sqrt{(k/k_0)^{2}-1}}\cos\left[-\frac\pi\xi+\frac{1}{\xi}\arctan \sqrt{(k/k_0)^{2}-1}\right]\Big\} ,  & k > k_0.
\end{array}
\right.
\end{equation*}

 Note that
 \[
   \lim_{k\to k_0-0}\omega_\text{r}(k) =-\tfrac{1}{2^{1/\xi}\tau}\sin\tfrac{\piup}{\xi}\,, \quad
   \lim_{k\to k_0+0}\omega_\text{r}(k)=\mp\tfrac{1}{2^{1/\xi}\tau}\sin\tfrac{\piup}{\xi}\,,
 \]
 \[
  \lim_{k\to k_0-0}\omega_\text{i}(k) =\lim_{k\to k_0+0}\omega_\text{i}(k) =\tfrac{1}{2^{1/\xi}\tau}\cos\tfrac{\piup}{\xi}\,,
 \]
 \[
   \lim_{k\to k_0-0}v_\text{p}(k)=-\tfrac{1}{2^{1/\xi}\tau k}\sin\tfrac{\piup}{\xi}\,, \quad
   \lim_{k\to k_0+0}v_\text{p}(k)=\mp\tfrac{1}{2^{1/\xi}\tau k}\sin\tfrac{\piup}{\xi}\,,
 \]
 i.e.,
 one branch of the real part of the frequency is continuous,
 the other branch has a first-order gap at $k=k_0$ (the same is for phase velocity),
whereas
 the imaginary part of the frequency is a continuous function everywhere.

 At $\xi=1$,
 we have well-known expressions for the ordinary Cattaneo-Maxwell equation:
 \begin{align*}
   \omega_\text{r}(k)=
   \left\{
     \begin{array}{ll}
       0, &                       0\leqslant k \leqslant k_0\,,\\
       \mp \frac1{2\tau}\sqrt{(k/k_0)^{2}-1},  & k > k_0\,,
     \end{array}
   \right.
   \quad
   \omega_\text{i}(k)=
   \left\{
     \begin{array}{ll}
       -\frac{{1\mp\sqrt{1-(k/k_0)^{2}}}}{2\tau}\,, &
                      0\leqslant k \leqslant k_0\,,\\[2mm]
       -\frac{1}{2\tau}\,,  & k > k_0\,,
     \end{array}
   \right.
\\
   v_\text{p}(k)=
   \left\{
     \begin{array}{ll}
       0, &                       0\leqslant k \leqslant k_0\,,\\
       \mp \frac1{2\tau k}\sqrt{(k/k_0)^{2}-1},  & k > k_0\,,
     \end{array}
   \right.
   \quad
   v_\text{g}(k)=
   \left\{
     \begin{array}{ll}
       0, &
                      0\leqslant k \leqslant k_0\,,\\
       \pm\frac{1}{2\tau}\frac{k/k_0^2}{\sqrt{(k/k_0)^2-1}}\,,  & k > k_0.
     \end{array}
   \right.
 \end{align*}\\[-7mm]

\begin{figure}[!t]
%  \vspace{-3mm}
  \centering
\includegraphics[width=0.80\textwidth]{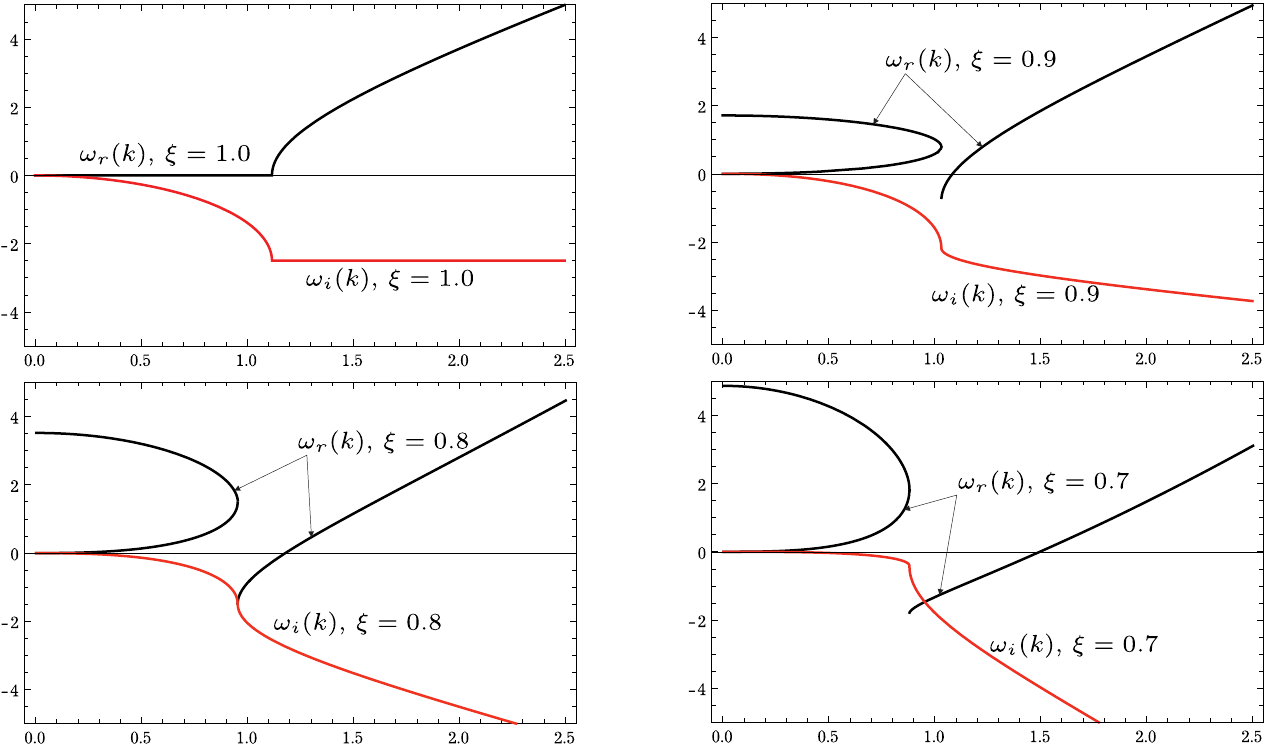}\\[-3mm]
  \caption{(Colour online) The frequency spectrum for $\tau=0.2$, $\bar{D}'=1$.\label{fig2}} \medskip
\end{figure}

\begin{figure}[!t]
  \centering
\includegraphics[width=0.80\textwidth]{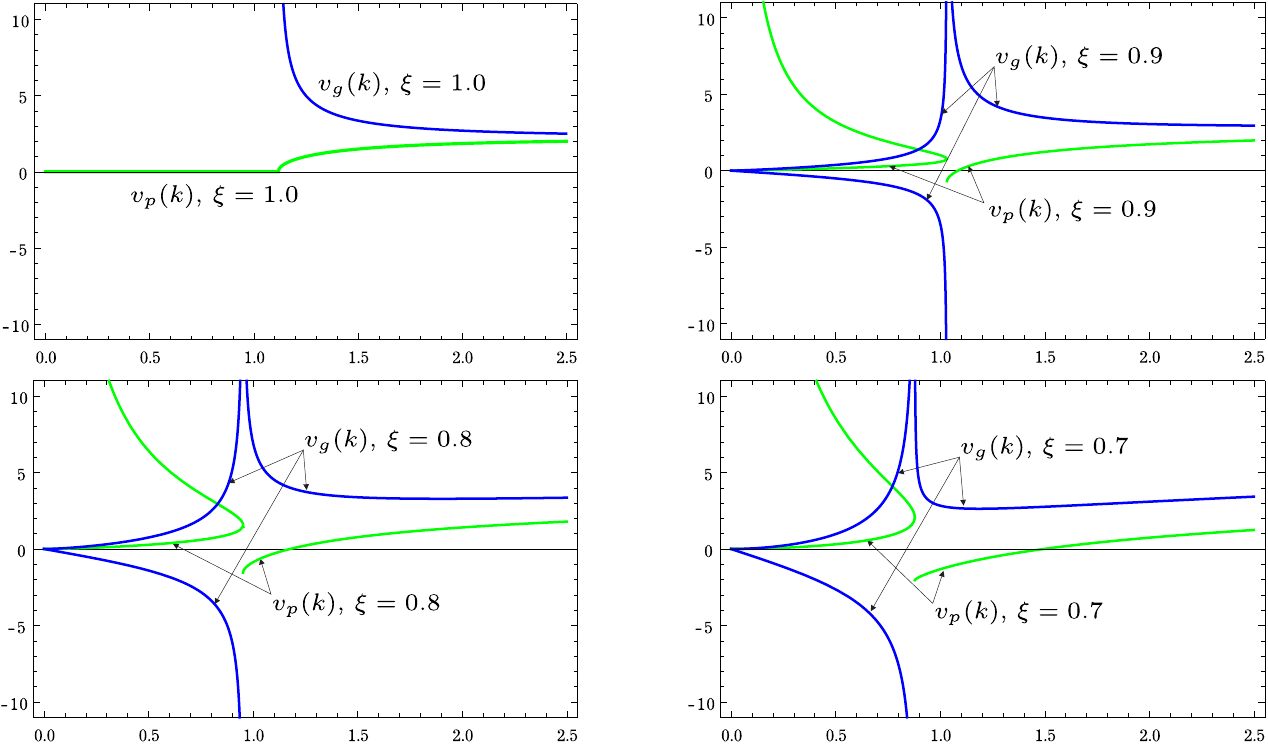}\\[-3mm]
  \caption{(Colour online) The phase $v_\text{p}(k)$ and group $v_\text{g}(k)$ velocities for $\tau=0.2$, $\bar{D}'=1$.\label{fig4}}
  \vspace{-4mm}
\end{figure}

 Using the analytic expressions for the frequency spectrum,
 numerical calculations are carried out at fixed values of the relaxation time $\tau=0.2$ and the diffusion coefficient $\bar{D}'=1$ with different values of the order $\xi$ of the fractional derivative, $\xi=1.0,\,0.9,\,0.8,\,0.7$.
 The results of the calculations are presented in figures~\ref{fig2}--\ref{fig4}.
 In the domain $k>k_0$,
 the signs are chosen so that the corresponding branches of the curves at $\xi=1$ coincide with the corresponding well-known results for the ordinary Cattaneo-Maxwell equation.
 Instead,
 there are two branches of the real part of the frequency in the domain $0\leqslant k \leqslant k_0$,
 as they both at $\xi=1$ coincide with the corresponding well-known results for the ordinary Cattaneo-Maxwell equation.
 Due to the nonlinearity of the equation,
 the point $k_0$ is a bifurcation point that disappears at $\xi=1$.
 At this point,
 two solutions appear,
 and when $\xi$ tends to $1$,
 they approach each other until a complete coincidence is reached.
 As the parameter $\xi$ decreases ($\xi=0.9,\,0.8,\,0.7$),
 the real part of the frequency $\omega_\text{r}(k)$ becomes a discontinuous function with an increasing gap at $k=k_0$,
 the imaginary part of the frequency $\omega_\text{i}(k)$ stops to be constant at $k>k_0$,
 $\omega_\text{i}(k)$ decreases.
 According to the frequency spectrum,
 the behaviour of the phase and group velocities changes with a decrease of $\xi$.
 For $\xi=0.9,\,0.8,\,0.7$ the phase velocity $v_\text{p}(k)$ has a gap,
 and the group velocity $v_\text{g}(k)$ has $\lambda$-like behaviour.
 Such a behaviour of the frequency spectrum,
 phase and group velocities in diffusion processes with corresponding characteristic relaxation time and diffusion coefficient may indicate a change in the properties of the environment in which the diffusion of particles takes place.
 These mechanisms are determined by the characteristic relaxation time~$\tau$ and the diffusion coefficient $\bar{D}'$,
 which resulted from the corresponding modelling of the memory function~(\ref{eq:2.194}).
 During the study of the frequency spectrum,
 the parameters $\tau$ and $\bar{D}'$ are fixed,
 we change the values of the fractional derivative index $\xi$.
 Obviously,
 for real diffusion processes
 (see, for example,~\cite{Kostrobij20184099}),
 the parameters $\tau$ and $\bar{D}'$ would be matched by a specific value of the fractional derivative index $\xi$,
 which would characterize the anomaly of the process.
 There is a transition from a normal diffusion process with characteristic relaxation time,
 described by the ordinary Cattaneo-Maxwell equation,
 to an anomalous mode,
 described by the Cattaneo-Maxwell equation with fractional derivatives,
 with corresponding discontinuities for the real part of the frequency spectrum,
 which describes the spread of the process,
 and the $\lambda$-like behaviour of the group velocities.
 This is a model research and it is obvious that features of the frequency spectrum studies would be important and quite interesting while investigating the diffusion wave processes for specific real systems.

 \section{Conclusions}
 We briefly presented the general approach for obtaining the generalized transport equations with the fractional derivatives by using the Liouville equation with the fractional derivatives for a system of classical particles and the Zubarev nonequilibrium statistical operator method within the Gibbs statistics.
 By using this approach,
 the model memory functions,
 and fractional calculus,
 the generalized Cattaneo-Maxwell diffusion equations with taking into account the space-time nonlocality have been obtained.

 The dispersion equation for the Cattaneo-Maxwell-type diffusion equation is found,
 taking into account the time-spatial nonlocality in fractional derivatives.
 In the case of the time nonlocality,
 solutions of this dispersion equation are found.
 It is found that the frequency spectrum,
 the corresponding phase, and group velocities have a bifurcation point,
 which disappears as the fractional derivative index $\xi$ tends to $1$.
 Besides,
 it is found that the real part of the frequency and the corresponding phase velocity have the first-order gap with a jump that increases with a decreasing parameter~$\xi$.
 Instead,
 the imaginary part of the frequency is a continuous function,
 and the group velocity has a singularity at the bifurcation point.

\ukrainianpart

\title{Узагальнене рівняння дифузії з просторово-часовою нелокальністю. Моделювання функції пам'яті}

\author{П.П. Костробій\refaddr{label1}, Б.М. Маркович\refaddr{label1}, М.В. Токарчук\refaddr{label2,label1}}

\addresses{
\addr{label1} Національний університет ``Львівська політехніка'', вул. С. Бандери, 12, 79013 Львів, Україна
\addr{label2} Інститут фізики конденсованих систем НАН України, вул. Свєнціцького, 1, 79011 Львів, Україна
}

\makeukrtitle

\begin{abstract}
\tolerance=3000%
   Представлено загальний підхід для отримання узагальнених рівнянь переносу з дробовими похідними,
   використовуючи рівняння Ліувілля з дробовими похідними для системи класичних частинок та метод нерівноважного статистичного оператора Зубарєва в статистиці Гіббса.
   Отримані нові немарковські рівняння дифузії іонів у просторово неоднорідному середовищі з фрактальною структурою та узагальненим рівнянням дифузії Каттанео-Максвелла з урахуванням просторово-часової нелокальності.
   Знайдено дисперсійні співвідношення для рівняння дифузії Каттанео-Максвелла з урахуванням просторово-часової нелокальності в дробових похідних.
   Розраховано частотний спектр, фазову та групову швидкості.
   Показано, що частотний спектр має хвильову поведінку з розривами,
   які також проявляються в поведінці фазової швидкості.
\keywords рівняння Каттанео, рівняння дифузії Каттанео-Максвелла, статистика Гіббса, нерівноважний статистичний оператор

\end{abstract}

\end{document}